\documentclass[times,twocolumn,final]{elsarticle}

\newcommand{\add}[1]{\textcolor{black}{#1}}

\newcommand{\addtable}[1]{{\color{black}#1}}

\newcommand{\vertices}{$\mathcal{V}$}
\newcommand{\edges}{$\mathcal{E}$}
\newcommand{\faces}{$\mathcal{F}$}

\newcommand{\customref}[2]{\hyperref[#1]{#2}}

\usepackage{siunitx}

\usepackage{medima}
\usepackage{framed,multirow}
\usepackage{subcaption}

\usepackage{amssymb}
\usepackage{latexsym}

\usepackage{url}
\usepackage[table]{xcolor}  

\usepackage{hyperref}
\usepackage{dblfloatfix}    
\usepackage{booktabs}
\usepackage{float}
\usepackage{lipsum}
\usepackage{graphicx}


\definecolor{newcolor}{rgb}{.8,.349,.1}

\journal{Medical Image Analysis}

\begin{document}

\verso{Rudolf van Herten \textit{et~al.}}

\begin{frontmatter}

\title{World of Forms: Deformable Geometric Templates for One-Shot Surface Meshing in Coronary CT Angiography}

\author[1,2,3]{Rudolf L.M. \snm{van Herten}\corref{cor1}}
\cortext[cor1]{Corresponding author}
\ead{r.l.m.vanherten@amsterdamumc.nl}
\author[1,2,3]{Ioannis \snm{Lagogiannis}}
\author[4]{Jelmer M. \snm{Wolterink}}
\author[1]{Steffen \snm{Bruns}}
\author[5]{Eva R. \snm{Meulendijks}}
\author[6]{Damini \snm{Dey}}
\author[3]{Joris R. \snm{de Groot}}
\author[3]{Jos\'{e} P. \snm{Henriques}}
\author[5,7]{R. Nils \snm{Planken}}
\author[1,2,3]{Simone \snm{Saitta}}
\author[1,2,3,5]{Ivana \snm{I\v{s}gum}}

\address[1]{Department of Biomedical Engineering and Physics, Amsterdam UMC, Meibergdreef 9, Amsterdam, 1105 AZ, The Netherlands}
\address[2]{Informatics Institute, University of Amsterdam, Amsterdam, The Netherlands}
\address[3]{Amsterdam Cardiovascular Sciences, Amsterdam, The Netherlands}
\address[4]{Department of Applied Mathematics, Technical Medical Centre, University of Twente, Drienerlolaan 5, Enschede, 7522 NB, The Netherlands}
\address[5]{Department of Radiology and Nuclear Medicine, Amsterdam UMC, Meibergdreef 9, Amsterdam, 1105 AZ, The Netherlands}
\address[6]{Biomedical Imaging Research Institute, Cedars-Sinai Medical Center, Los Angeles, USA}
\address[7]{Department of Radiology, Mayo Clinic, Rochester, USA}

\received{-}
\finalform{-}
\accepted{-}
\availableonline{-}
\communicated{-}

\begin{abstract}
Deep learning-based medical image segmentation and surface mesh generation typically involve a sequential pipeline from image to segmentation to meshes, often requiring large training datasets while making limited use of prior geometric knowledge. This may lead to topological inconsistencies and suboptimal performance in low-data regimes. To address these challenges, we propose a data-efficient deep learning method for direct 3D anatomical object surface meshing using geometric priors. Our approach employs a multi-resolution graph neural network that operates on a prior geometric template which is deformed to fit object boundaries of interest. We show how different templates may be used for the different surface meshing targets, and introduce a novel masked autoencoder pretraining strategy for 3D spherical data. The proposed method outperforms nnUNet in a one-shot setting for segmentation of the pericardium, left ventricle (LV) cavity and the LV myocardium. Similarly, the method outperforms other lumen segmentation operating on multi-planar reformatted images. Results further indicate that mesh quality is on par with or improves upon marching cubes post-processing of voxel mask predictions, while remaining flexible in the choice of mesh triangulation prior, thus paving the way for more accurate and topologically consistent 3D medical object surface meshing.
\end{abstract}


\end{frontmatter}

\section{Introduction}
\label{sec:introduction}
Many applications in 3D cardiovascular image analysis require image segmentation to produce manifold surface meshes. This allows one to perform functional analyses of tissues at different scales, for which examples include computational fluid dynamics~\citep{suk2021mesh}, electrophysiological modeling~\citep{cedilnik2019fully}, and cardiac motion and strain analysis~\citep{szilveszter2020left}. A common first step towards generating these manifolds involves the automatic voxelwise segmentation of the structure of interest. This step is typically performed by training a deep neural network to perform dense 3D voxelwise classification on the input grid, for which both convolutional- and transformer-based architectures perform effectively~\citep{isensee2021nnu, hatamizadeh2022unetr}. However, voxelwise classification implicitly introduces various limitations. First, manual labeling of medical images to obtain reference segmentations for training and evaluation of automatic methods is considered a labor-intensive task that may require a high level of domain-specific expertise. Second, generated voxel masks discretize the underlying anatomical shape, and are thus limited to the resolution of the input grid. Furthermore, masks are neither guaranteed to be contiguous nor free of anatomically implausible object delineations~\citep{kong2021deep}. As such, generating high-quality meshes from grid-based methods through typical marching cubes approaches~\citep{lorensen1998marching} is error-prone and requires ad hoc post-processing.

To address these issues, an increasing number of works have focused on the efficient generation of continuous surface meshes from discrete image inputs~\citep{bohlender2021survey}. A common approach to explicitly include shape information in deep learning is to predict a parameterization of the desired shape rather than a voxel mask~\citep{attar20193d}. This can be done at a global level by describing the structure through a set of shape descriptors~\citep{kervadec2021beyond}, or locally, by generating a high-resolution parameterization of the shape. For example, \cite{alblas2022going} propose the use of implicit neural representations to obtain continuous implicit shape representations from existing explicit ones in 3D vascular modeling. Another option is to infer triangulated mesh representations with \textit{a priori} fixed connectivity directly from image data. In this approach, a displacement vector is predicted for each vertex in a template mesh based on local image information~\citep{wickramasinghe2020voxel2mesh, kong2021deep}. A challenge to this strategy is its susceptibility to self-intersecting mesh faces and the generation of low-quality meshes with small angles and high aspect ratios. To prevent faces from intersecting, regularization of the mesh Laplacian, normals, and edge lengths is required in e.g. \citep{kong2021deep}. However, many shapes in the human body are \textit{star-convex}, i.e. a set of points exists from which the entire shape boundary is visible. For these shapes, it might be sufficient to limit displacement of mesh vertices to straight lines or \textit{rays} in 3D that are cast from a central seed point from within the object boundary. Conversely, more complex shapes such as arteries can be reduced to a trace or \textit{centerline} of the object, for which the boundary can be modeled by casting rays in tangent planes along the centerline.

In conventional image analysis, 2D ray casting has previously been employed for femur and tibia segmentation in orthogonal planes in knee MRI~\citep{dodin2011fully}, as well as segmentation of epicardial adipose tissue in cardiac CT~\citep{barbosa2011towards}. In 3D, convex segmentations have been obtained of abdominal aortic aneurysms, or concave segmentations of kidneys by combining two convex meshes~\citep{kronman2012anatomical}. However, in these works, displacement was determined based on hand-crafted rules and gradient thresholds, leading to errors that required post-processing such as Laplacian spherical interpolation. As an alternative, learning-based ray casting was proposed in multiple works for coronary artery segmentation with the assumption of a cylindrical shape~\citep{lugauer2014precise, van2023automatic}. Similarly, prior knowledge about the cylindrical shape of carotid arteries was used for segmentation in 3D polar coordinate systems~\citep{alblas2022deep}. Finally, \cite{wolterink2019graph} used graph convolutional networks (GCN)~\citep{kipf2016semi} to combine information along equiangular rays orthogonal to coronary centerlines with that of neighboring rays to obtain a smooth coronary artery surface mesh.

\begin{figure}[t]
    \centering
    \includegraphics[trim={0.5cm 1cm 1cm 1.75cm}, clip, width=\columnwidth]{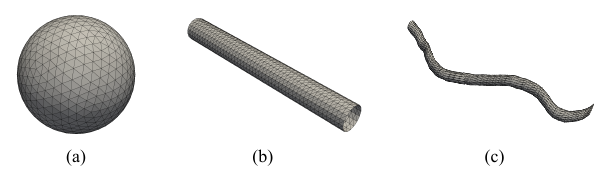}
    \caption{Geometric priors explored in this work. (a) The sphere $\mathbb{S}^2$, for 3D star-convex shapes, (b) the cylinder, for modeling shapes as straight tubes, and (c) the tube in 3D space, implicitly containing prior curvature information.}
    \label{fig:shapes}
\end{figure}

Traditional voxel-based methods typically require a substantial amount of fully-labeled data to perform well on unseen samples, with manual annotation potentially taking multiple hours per case~\citep{bruns2022deep}. This is especially problematic in the medical imaging domain, where annotated data is typically scarce and costly to obtain. Incorporating shape priors therefore provides an effective way to limit the solution space while implicitly producing continuous surface meshes. Specifically, this constrained solution space presents a unique opportunity for one-shot learning, where models can potentially generalize from a single labeled example. 
In this work, we present a weakly-supervised, one-shot method for the automatic generation of smooth surface meshes from prior-informed geometric representations of 3D medical images (see Fig.~\ref{fig:shapes}). Geometric representations are generated by casting rays from a seed point within the structure of interest, sampled from the input image. This representation is subsequently processed by a \textit{mesh neural network} that directly operates on the geometric prior while retaining desirable shape properties such as rotation equivariance along the angular dimension in the case of a cylindrical shape prior. Furthermore, for the specific case of learning on the sphere, we propose a pretraining strategy inspired by masked autoencoders (MAE)~\citep{he2022masked}, further enhancing effectiveness in the low data regime. We evaluate the proposed method on three different segmentation tasks in cardiac CT angiography (CCTA). The first task concerns the segmentation of the intra-pericardial space, which can be used for the quantification of epicardial adipose tissue~\citep{commandeur2018deep}. Secondly, we evaluate performance on segmentation of the left ventricle (LV) cavity and the LV myocardium, and finally, we evaluate performance on coronary artery lumen segmentation.

The remainder of this manuscript is structured as follows: Section~\ref{sec:data} describes the dataset. Section~\ref{sec:method} describes the methodology, for which experimental details are covered in Section~\ref{sec:exp}. A comparison with baseline methods and ablations is provided in Section~\ref{sec:res}, and results are discussed in Section~\ref{sec:disc}.

\begin{figure*}[t]
\centering
     \includegraphics[width=1.\textwidth]{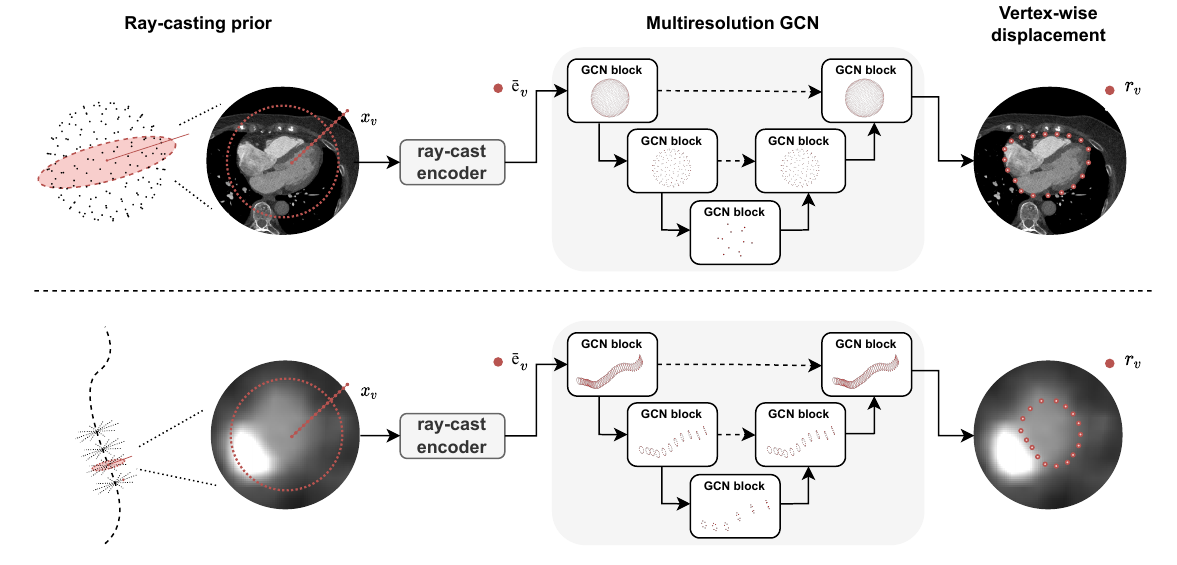}
      \caption{Overview of the proposed method. A geometric prior is chosen to represent an object of interest, on which image information is projected by casting rays from a central point or line within the object of interest. An encoder subsequently embeds each ray cast $x_v$ to a latent vertex representation $\overline{e}_v$. A template mesh describing the geometric prior is used to process these embeddings using a multiresolution GCN, allowing the network to incorporate both local and global image information. For each vertex, the network regresses the distance to the object boundary. The top row displays the use of a spherical prior for defining the pericardium mesh, while the bottom row displays the use of a coronary artery centerline prior for coronary artery lumen meshing.}
       \label{overview}
\end{figure*}

\section{Data}

\label{sec:data}
This study comprises two datasets of retrospectively collected CCTA scans obtained from clinical routine, as well as two public datasets. The two clinical datasets originate from the Amsterdam University Medical Center, The Netherlands (Set 1 and 2). The two public datasets contain a subset of the Multi-Modality Whole Heart Segmentation (MM-WHS) challenge dataset (Set 3)~\citep{zhuang2019evaluation}, and the training dataset from the Automated Segmentation of Coronary Arteries (ASOCA) challenge (Set 4)~\citep{gharleghi2023annotated}.

Set 1 contains CCTA exams of 394 patients, acquired on a Siemens Somatom Force CT scanner (Siemens Healthineers, Erlangen, Germany). All patients underwent the workup for transcatheter aortic valve implantation. The median age was 80 years (range 36--93 years) and 192 patients were male. Tube voltage ranged from 70 to 120 kVp and the tube current ranged from 296 to 644 mAs. Scans were reconstructed with an in-plane resolution of 0.26-0.46 mm and a slice thickness and increment of 0.6 mm.

Set 2 features 148 CCTA images from 74 patients undergoing thoracoscopic pulmonary vein antrum isolation~\citep{krul2011thoracoscopic}. The median age was 63 years (range 37--78 years) and 52 patients were male. For each patient, one CCTA scan was acquired prior to intervention and one was acquired at six-month follow-up to determine pulmonary vein stenosis. 102 CCTA scans were acquired on a Somatom Force CT scanner (Siemens Healthineers, Erlangen, Germany), 32 scans on a Somatom Definition AS+ (Siemens Healthineers, Erlangen, Germany), and 14 scans on a Sensation 64 (Siemens Healthineers, Erlangen, Germany). The median tube voltage was 95 kVp (range 80--120 kVp) and the median tube current was 225 mAs (range 41--598 mAs). Images were reconstructed to a median in-plane isotropic voxel size of 0.68 mm (range 0.35--0.86 mm) with a median slice thickness and increment of 1.0~mm (range 0.6--1.5 mm).

Set 3 includes the CCTA training dataset of the MM-WHS challenge, which features 20 end-diastolic images~\citep{zhuang2019evaluation}. These images have a median in-plane voxel size of 0.44 mm (range 0.28--0.59 mm) with a mean slice thickness of 1.6~mm.

Set 4 features the CCTA scans from the ASOCA training dataset, which includes 20 healthy subjects and 20 patients with confirmed coronary artery disease~\citep{gharleghi2023annotated}. Images have a median in-plane voxel size of 0.40 mm (range 0.31--0.49 mm) and a mean slice thickness of 0.6~mm.

\section{Method}
\label{sec:method}

We present a method for the direct prediction of surface meshes using geometric priors, which are induced through knowledge regarding the global shape of objects of interest. This prior is described by a deformable contiguous mesh $\mathcal{M}$~$({\mathcal{V}, \mathcal{E}, \mathcal{F}})$, with vertices \vertices{}, edges \edges{}, and faces \faces{}. Each vertex location on the mesh is characterized by an intensity profile along a ray cast from a central seed point.
The goal is to define a neural network leveraging image information along with the innate mesh definition to displace the vertices along ray-casts to corresponding location on the object boundary, i.e. to define the surface of the object of interest. Therefore, we propose a two-stage approach. In the first stage, an embedding network encodes ray-cast image information for each vertex on the mesh. In the second stage, embeddings are processed by a multiresolution \textit{mesh neural network}, which leverages relative vertex positions to perform anisotropic convolutions along edges \edges{} at multiple resolutions on a manifold of choice. This allows the network to learn local and global information relevant for accurate delineation of the object of interest. A general overview of the method is presented in Fig.~\ref{overview}.

\subsection{Geometric priors}
\label{sec:method priors}
\begin{figure}[!b]
    \centering
    \scalebox{-1}[1]{\includegraphics[trim={0cm 0.9cm 0cm 0.45cm}, clip, width=\columnwidth]{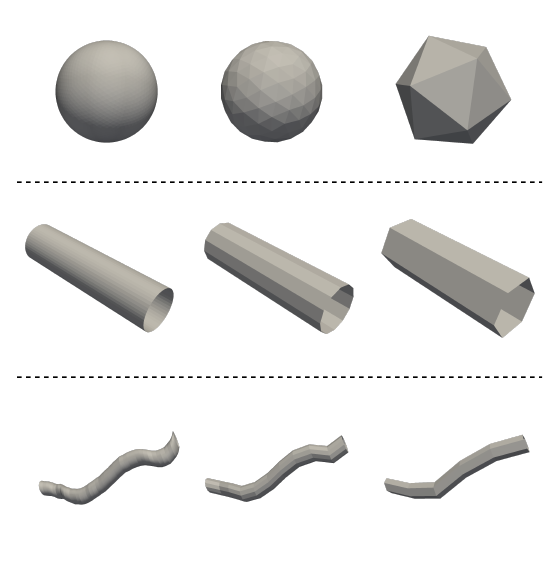}}
    \caption{Geometric priors at different resolutions. The top row displays the icosahedron at 0, 2, and 4 subdivision steps. The middle and bottom rows display the cylinder and the tube deformed in 3D space respectively, at 1, 2, and 4 subdivision steps.}
    \label{fig:scales}
\end{figure}

Since the template shape of many cardiac structures can be reduced to either a sphere (e.g. the LV myocardium) or a tube (e.g. coronary arteries), these geometric objects are considered for our task of prior-informed mesh generation. Prior meshes are therefore generated to discretize each geometry (see Fig.~\ref{fig:scales}).

To approximate uniformly distributed vertices, we model the sphere as a subdivided icosahedron (i.e. an~\textit{icosphere}). A regular icosahedron is composed of 20 triangular faces defined by 30 edges, with 12 vertices lying on its circumscribed sphere. In each subdivision step, every face is replaced by four new faces with all vertices lying on the circumscribed sphere. This is achieved by adding one new vertex per icosahedron edge. The new vertex is projected from the center of the old edge onto the circumscribed sphere. The new vertices are connected with new edges such that for each triangular face in the previous configuration, four new triangular faces are created. The number of subdivision steps can be varied arbitrarily, allowing us to vary the number of vertices, edges, and faces, and thus the resolution of the template mesh.

The cylinder may be discretized similarly by defining cross-sections (i.e. a circle) as equiangular triangles. This coarse representation can subsequently be subdivided to a higher resolution by adding a new vertex at the center of each edge, and projecting its location back onto the cylinder. Triangular faces are defined by connecting vertices in adjacent cross-sections at arbitrary resolution as defined in \cite{wolterink2019graph}.

Finally, given a centerline prior, the cylinder may be deformed by defining cross-sections at planes orthogonal to the centerline direction. This results in a tube deformed in 3D space, for which curvature is implicitly embedded in the vertex normals.

\begin{figure*}[h]
\centering
     \includegraphics[width=1.\textwidth]{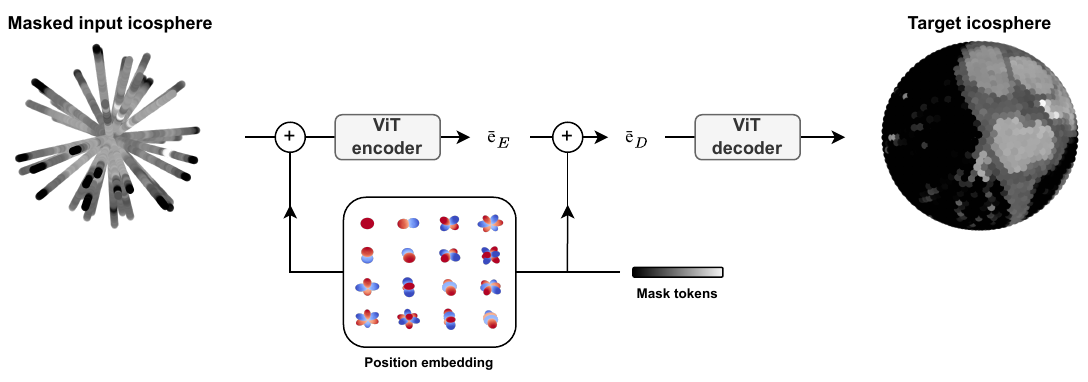}
      \caption{Pretraining on the icosphere using a spherical masked autoencoder. Each ray-cast is treated as a token, for which relative position embeddings are generated using spherical harmonics descriptors. Similar to the original masked autoencoder, a vision transformer (ViT) encodes the input using a small set of unmasked ray-cast tokens. Subsequently, mask tokens and their positional embeddings are appended to the encodings $\overline{e}$ and simultaneously decoded to reconstruct all ray-casts in the target icosphere. At test-time, this allows the ViT encoder to embed global information into each ray-cast representation using the full unmasked image volume.}
       \label{fig:MAE}
\end{figure*}

\subsubsection{Ray casting}
\label{sec:method raycasting}
Local image information is embedded for each vertex $v$ on template meshes through ray casting. A \textit{ray} is defined as all points laying on a line that starts from some origin location $c$ and intersects $v$. By sampling the image intensity values at points along this line, a 1D signal $x_v$ is obtained that represents the image information along the ray cast. For each vertex, a total of $n_v$ equidistant points are sampled with spacing $s_v$ such that the ray is sufficiently large to escape the structure of interest (see Fig.~\ref{overview}).

The location of each vertex on the spherical template mesh resulting from icosahedron subdivision can be described in spherical coordinates given an origin or seed point $c$ that lies within the mesh. That is, for each vertex $v$, there is a polar angle $\phi_v$, an azimuthal angle $\theta_v$, and a radial distance $r_v$. Rays are cast for a fixed $\phi$ and $\theta$, and we aim to identify a value $r$ for each vertex such that the vertex lies on the boundary of the structure of interest. \add{By fixing the angles and only predicting the radial distance, the complex 3D surface generation task is reduced to a simpler 1D regression problem along each ray cast.}

Similarly, cylindrical coordinates may describe vertex locations on tubular structures. In this scenario, each ray cast is described by a location along a centerline $c_v$, and a polar angle $\phi_v$ in an orthogonal plane. The centerline is equivalent to the $z$-direction for a perfect cylinder, but may also describe a line curving in 3D space.

\subsection{Ray-cast encoding}

Since ray-casts are 1D vectors directly interpolated from the input image volume, they can be embedded into a latent vector describing the local image information.

\subsubsection{Convolutional encoder}

To process the image signal along a ray-cast, a 1D convolutional encoder may be employed, as it introduces a useful and lightweight bias while allowing for the extraction of descriptive features. Specifically, we employ ResNet skip-connection blocks for the convolutional encoder, and introduce a pooling layer after each block~\citep{he2016deep}. The number of skip-connection blocks is varied given the complexity of the input data, and is scaled with input signal size $n_v$ (see Appendix \customref{sec:conv encoder}{A.1}). 

\subsubsection{Transformer encoder: the icosphere MAE}
\label{sec:method pretraining}
For the special case where our geometric prior is a sphere, we may leverage MAE pretraining to produce informative ray-cast embeddings~\citep{he2022masked}. Similar to the original work, this network is tasked with recovering a complete icosphere image volume given only a small number of unmasked ray-casts in the input (see Fig.~\ref{fig:MAE}). The set of unmasked ray-casts is first embedded to latent vectors or \textit{tokens} through a linear layer. For each token, a position embedding is generated using the Fourier series for the sphere, i.e. spherical harmonics~\citep{e3nn_paper}. A vision transformer (ViT) encoder subsequently processes the set of unmasked tokens to embeddings $\overline{e}_E$. After adding mask tokens and position embeddings for the missing ray-casts (resulting in $\overline{e}_D$), a lightweight ViT decoder reconstructs the full image volume. As such, the ViT encoder learns to embed local and global semantic information for each ray-cast token. The decoder is discarded at test time, and embeddings are generated for the full set of ray-casts in the input volume. 

\subsection{Learning manifold vertex locations}
\label{sec:method radius prediction}
In order to predict $r_v$ for each vertex $v$ on the prior mesh $\mathcal{M}$, a neural network needs to be defined that can process $x_v$ and the local neighborhood of $v$. For example, representing arteries curving in 3D space as polar-transformed multi-planar reformatted images allows one to utilize convolutional neural networks (CNNs) on such warped images~\citep{alblas2022deep, van2023automatic}. However, any domain that cannot be treated as a regular grid does not allow for the use of CNNs, as convolutions are not well-defined for spheres or general meshes. Processing a ray-cast $x_v$ for arbitrary manifold configurations therefore requires the use of a domain-agnostic neural network. A GCN can therefore be used, which allows users to manually define the connectivity of a graph $\mathcal{G}$ (\vertices{}, \edges{}). A single network layer aggregates information of connected vertices through message passing~\citep{gilmer2017neural}: 

\begin{equation}
\label{eq1}
\mathbf{x}_v^{\prime} = \sigma \left( \bigoplus_{n \in \mathcal{N}(v)} \, f
        \left(\mathbf{x}_n\right) \right)
\end{equation}

where $v \in \mathcal{V}$ with neighbors $\mathcal{N}(v)$, $x_n$ denotes the hidden state of a neighboring vertex $n \in \mathcal{N}(v)$, and self-loops are typically added such that $v \in \mathcal{N}(v)$. The trainable layer $f(\cdot)$ describes the message passing interaction from $n$ to $v$, which is subsequently aggregated for all neighboring vertices on the central node $v$. Finally, $\sigma$ is a non-linear activation function that returns the updated vertex $x_v^{\prime}$ after aggregation. A network consisting of multiple GCN layers may then be used to approximate $r_v$.

A limitation of the standard GCN~\citep{kipf2016semi} is that, unlike CNNs, they are unaware of the relative positions of vertices, as they are typically designed as both isotropic and permutation invariant. To improve the expressivity of GCNs, graph attention networks (GATs) were introduced, which include a self-attention mechanism that weighs the importance of neighboring vertices based on their features~\citep{velivckovic2017graph, brody2021attentive}:

\begin{figure*}[t]
    \centering
    \includegraphics[trim={0cm 0cm 0cm 0cm}, clip, width=1.\textwidth]{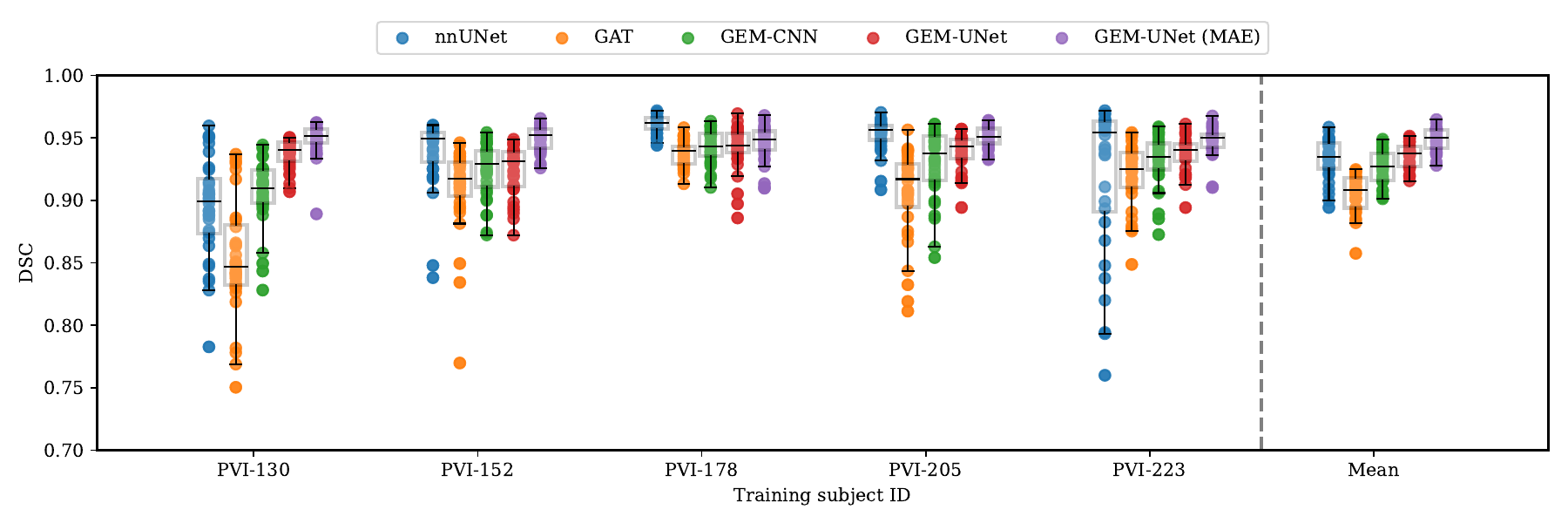}
    \caption{Test-set Dice similarity coefficient (DSC) for one-shot pericardium boundary segmentation across different models. The performance of nnUNet, GAT, GEM-CNN, GEM-UNet, and GEM-UNet with MAE embeddings is compared. Each model is trained for 5 training subject IDs, and the rightmost column shows the mean DSC aggregated over all patients for each model.}
    \label{fig:peri_dsc}
\end{figure*}

\begin{equation}
\label{eq1}
\mathbf{x}_v^{\prime} = \sigma \left(\bigoplus_{n \in \mathcal{N}(v)} \, g \left(\mathbf{x}_n, \mathbf{x}_v\right) f
        \left(\mathbf{x}_n\right) \right)
\end{equation}

where $g(\cdot)$ is a trainable function that returns an attention coefficient for the message from $n$ to $v$. Though the GAT layer is an anisotropic operator, note that it remains permutation invariant and does not consider the inherent local geometry of mesh data.

\subsubsection{Gauge-equivariant mesh convolutions}
Given these limitations, it is beneficial to employ a network that is implicitly sensitive to the relative positions of mesh vertices. Gauge equivariant mesh (GEM)-CNNs solve the issue of permutation invariance by introducing weights that are influenced by the relative orientation of vertex neighbors~\citep{dehaan2021}. Specifically, the network is designed to be equivariant to a local change of reference neighbor or \textit{gauge}, which ensures that the anisotropic convolutional operations remain consistent regardless of the choice of reference framework. This is achieved by incorporating parallel transport to align features before convolution:

\begin{equation}
\label{eq1}
\mathbf{x}_v^{\prime} = \sigma \left(\bigoplus_{n \in \mathcal{N}(v)} \, f \left(\theta_{v,n}, \rho_{x_n \rightarrow x_v} \mathbf{x}_n \right)\right)
\end{equation}

\begin{table}[b!]
\centering
\caption{Full test-set results for one-shot pericardium segmentation. Scores are averaged for five different training patient IDs. The bottom row represents an upper-bound for test-set performance in terms of a 5-fold nnUNet trained on the full training dataset. The Dice similarity coefficient (DSC), average symmetric surface distance (ASSD) and 95\% Hausdorff distance are presented. MAE indicates the use of masked autoenocoder-pretrained embeddings. Results are given as mean (standard deviation).}
\label{tab:peri_performance}
\resizebox{\columnwidth}{!}{
\begin{tabular}{lcccc}
\toprule
& Remarks & DSC $\uparrow$ & ASSD $\downarrow$ (mm) & HD95 $\downarrow$ (mm) \\
\midrule
\href{https://www.nature.com/articles/s41592-020-01008-z}{nnUNet} & & 0.933 (0.016) & 4.505 (3.405) & 26.776 (23.904) \\
\add{\href{https://link.springer.com/chapter/10.1007/978-3-031-72114-4_47}{nnUNet}} & \add{ResEnc-L} & \add{0.933 (0.023)} & \add{5.030 (4.311)} & \add{25.689 (20.337)} \\
\add{\href{https://link.springer.com/chapter/10.1007/978-3-031-43901-8_39}{MedNeXt}} & \add{L k3} & \add{0.923 (0.035)} & \add{4.705 (2.877)} & \add{22.484 (13.523)} \\
\midrule
GAT & & 0.905 (0.015) & 3.896 (0.748) & 17.635 (3.317) \\
GEM-CNN & & 0.927 (0.014) & 3.021 (0.680) & 11.107 (3.563) \\
GEM-UNet & & 0.936 (0.010) & 2.702 (0.575) & 9.603 (4.071) \\
GEM-UNet & MAE & \textbf{0.948} (0.009) & \textbf{2.174} (0.554) & \textbf{7.924} (4.113) \\
\midrule
\href{https://www.nature.com/articles/s41592-020-01008-z}{nnUNet} & & 0.973 (0.005) & 1.505 (1.522) & 4.655 (4.076) \\
\bottomrule
\end{tabular}
}
\end{table}

where $\rho$ is the parallel transport operator that aligns features from $x_n$ to $x_v$, and message passing $f(\cdot)$ is also a function of the relative angle $\theta$ between $v$ and $n$, defined on the tangent plane of $v$.

A benefit of GEM-CNNs is their ability to maintain equivariance under specific transformations given certain mesh priors. Specifically, if the mesh structure is preserved when mapping between vertices, the method becomes equivariant to the mapping transformation. For a perfect cylinder, this implies SO(2) equivariance along the angular dimension and translation equivariance along the $z$-dimension. Similarly, GEM-CNNs achieve SO(3) equivariance when applied to spherical data~\citep{alblas2023sire}.

\subsubsection{Multiresolution GCN}
Similar to regular CNNs, GCNs process the local neighborhood of vertices. Networks may however benefit from a larger context window for accurate surface meshing, as is the case with voxel-wise segmentation~\citep{ronneberger2015u}. \add{To address this, we leverage the hierarchical subdivision structure of our geometric priors described in Section \ref{sec:method priors}, where each subdivision step creates new vertices and faces from the previous level. This natural hierarchy allows us to create a multiresolution graph~(see Fig. \ref{fig:scales}) by connecting vertices across subdivision levels. For example, with an icosphere prior, vertices from the original icosahedron (subdivision level 0) are used as aggregation points for vertices at subdivision level 2.} For GEM-CNNs, parallel transport is necessary to average and redistribute node features at pooling and unpooling respectively~\citep{suk2024mesh}. Given how this network incorporates a UNet-like structure, we will hereafter refer to this network as \textit{GEM-UNet}.

\section{Experiments}
\label{sec:exp}
The effectiveness of the proposed method is tested for three segmentation tasks. The method is provided with a different geometric prior corresponding to the cardiac structure of interest for each segmentation task.

\add{All experiments, including baselines and ablation studies, are repeated five times, each time using a random different single training sample to assess one-shot performance, effectively creating a five-fold cross-validation for one-shot learning.}

\subsection{Spherical prior}
\label{sec: exp spherical}
For all spherical experiments, the multiresolution GCN operates on icosphere subdivision levels of 4, 2, and 0 from highest to lowest resolution, respectively. All mesh vertices feature a corresponding ray-cast with a length of 128 mm, sampled at a resolution of 0.5 mm to ensure that all rays escape the object of interest. We consider the segmentation of the intra-pericardial space, and the segmentation of two nested surfaces: the LV cavity and the LV myocardium. These structures may be used to quantify epicardial adipose tissue or the LV ejection fraction. 

\begin{table*}[t]
\centering
\caption{Quantitative scores for different model versions on one-shot left ventricle (LV) cavity and myocardium (Myo) segmentation in the MM-WHS dataset. Scores are averaged for five different training images. The bottom row represents an upper-bound for test-set performance in terms of a 5-fold nnUNet trained on the full training dataset. Dice similarity coefficient (DSC), average symmetric surface distance (ASSD), and 95\% Hausdorff distance (HD95) metrics are reported. Results are presented as mean (standard deviation). The MAE column indicates use of masked autoencoder tokens rather than a 1D-convolutional encoder.}
\label{tab:model_performance}
\resizebox{0.85\textwidth}{!}{
\begin{tabular}{lccccccc}
\toprule
& Remarks & \multicolumn{2}{c}{DSC $\uparrow$} & \multicolumn{2}{c}{ASSD $\downarrow$ (mm)} & \multicolumn{2}{c}{HD95 $\downarrow$ (mm)}\\
\cmidrule(lr){3-4} \cmidrule(lr){5-6} \cmidrule(lr){7-8}
& & LV & Myo & LV & Myo & LV & Myo \\
\midrule
\href{https://www.nature.com/articles/s41592-020-01008-z}{nnUNet} & & 0.893 (0.055) & 0.849 (0.054) & 4.206 (4.453) & 4.254 (4.884) & 17.036 (15.058) & 20.219 (18.362) \\
\add{\href{https://link.springer.com/chapter/10.1007/978-3-031-72114-4_47}{nnUNet}} & \add{ResEnc-L} & \add{0.900 (0.068)} & \add{0.817 (0.094)} & \add{2.381 (1.737)} & \add{3.699 (2.595)} & \add{13.662 (12.435)} & \add{24.313 (18.929)} \\
\add{\href{https://link.springer.com/chapter/10.1007/978-3-031-43901-8_39}{MedNeXt}} & \add{L k3} & \add{0.813 (0.102)} & \add{0.799 (0.059)} & \add{3.596 (2.877)} & \add{3.587 (3.136)} & \add{13.034 (10.665)} & \add{22.357 (19.624)} \\
\midrule
GAT & & 0.780 (0.081) & 0.694 (0.088) & 3.676 (1.168) & 3.089 (0.715) & 17.015 (3.164) & 17.545 (4.033) \\
GAT & MAE & 0.896 (0.044) & 0.858 (0.044) & 1.457 (0.385) & 1.163 (0.222) & 5.713 (1.501) & 4.043 (1.191) \\
GEM-CNN & & 0.809 (0.068) & 0.707 (0.074) & 3.098 (0.992) & 2.569 (0.387) & 12.500 (2.634) & 11.311 (1.596) \\
GEM-CNN & MAE & \textbf{0.910} (0.034) & \textbf{0.862} (0.049) & \textbf{1.235} (0.261) & \textbf{1.108} (0.248) & \textbf{3.786} (0.874) & \textbf{3.253} (0.846) \\
GEM-UNet & & 0.861 (0.060) & 0.654 (0.114) & 1.954 (0.600) & 2.934 (1.294) & 7.058 (2.756) & 7.818 (2.839) \\
GEM-UNet & MAE & 0.907 (0.041) & 0.799 (0.070) & 1.270 (0.372) & 1.665 (0.728) & 3.812 (1.207) & 4.114 (1.700) \\
\midrule
\href{https://www.nature.com/articles/s41592-020-01008-z}{nnUNet} & & 0.930 (0.037) & 0.903 (0.017) & 0.956 (0.803) & 0.803 (0.179) & 2.652 (1.049) & 2.539 (0.705) \\
\bottomrule
\end{tabular}
}
\end{table*}

\subsubsection{Pre-training}
\label{sec: exp pre-training}
Since delineation of the pericardium, the LV cavity, and the LV myocardium all rely on 3D CCTA input volumes, we consider collective icosphere pre-training for all involved tasks (see Section \ref{sec:method pretraining}). For pre-training, an icosphere prior is generated at the 4th subdivision step, resulting in a mesh with 2,562 vertices. Rays are cast from a central seed point within the pericardium boundary, which is augmented by randomly offsetting the seed by up to 40 mm in all directions. We randomly mask out 95\% of the vertices during training to ensure that the MAE learns an effective ray-cast embedding. Once trained, the MAE weights are frozen for downstream segmentation tasks. For the full training and implementation details, we refer to Appendix~\customref{sec:masked autoencoder}{A.2}. 

\textit{Data:} The pre-training dataset used in this work was derived from Set~1, from which a total of 382 patients were selected. For each patient, both an end-systolic and end-diastolic CCTA scan were obtained, resulting in a total of 764 images. For each scan, a single seed point was annotated in a central location of the heart.

\subsubsection{Intra-pericardial space segmentation}
\label{sec: exp pericardium}
To assess surface meshing performance, we apply the proposed method to the delineation of the pericardial boundary. The network is optimized by minimizing the mean absolute error between predicted vertex radii $\hat{r}_v$ and the reference radii $r_v$. 

\begin{equation}
\add{\mathcal{L}_{peri} = \frac{1}{K}\sum_{v \in K} |r_v - \hat{r}_v|,}
\end{equation}

\add{where $K$ describes the set of all vertices in a batch.} Full optimization details are provided in Appendix~\customref{sec:gcn details}{A.3}.

\textit{Data:} For segmentation of the intra-pericardial space, all available CCTA exams from Set~2 were used. Of the 74 patients in this set, 16 patients were used for testing, while the remaining 58 patients were \add{used for one-shot cross-validation}. Reference segmentations of the intra-pericardial space were obtained in all 148 CCTA images. A semi-automatic research tool (QFAT, Cedars Sinai, Los Angeles, USA) was used to visualize the CCTA image and manually set markers on the pericardial boundary in axial slices. The cranial boundary was defined as the splitting of the pulmonary arteries, and the distal boundary was the apex of the heart. The exported markers were connected per axial slice through B-spline interpolation, and the resulting 2D contour was filled to obtain a 2D mask of the intra-pericardial space. Between the axial slices in which markers were set, a shape-based interpolation between the 2D masks was performed to obtain a 3D reference mask of the intra-pericardial space.

\begin{figure}[b!]
    \centering
    \includegraphics[trim={0.5cm 0.25cm 0.5cm 0cm}, clip, width=\columnwidth]{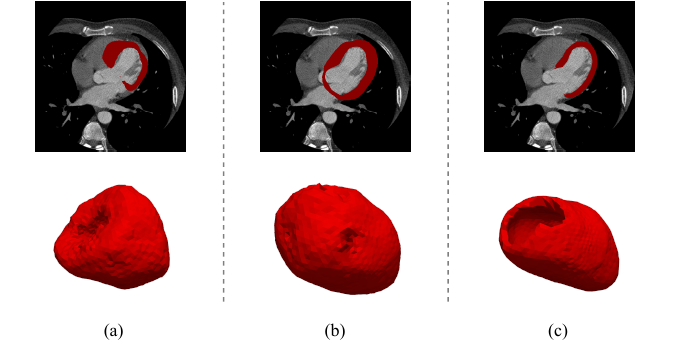}
    \caption{LV myocardium meshes for a test-set image from the MM-WHS dataset, generated using different one-shot models. (a) shows the result for the GEM-CNN, (b) utilizes the GEM-UNet, and (c) uses a GEM-CNN with MAE tokens as ray-cast embeddings. While (b) correctly identifies most of the inner myocardial boundary, only (c) is able to accurately delineate both the inner and outer myocardial boundary.}
    \label{fig:myo_comparison}
\end{figure}

\subsubsection{Left ventricle cavity and myocardium segmentation}
\label{sec: exp myocardium}
For the delineation of the nested LV cavity and LV myocardium, two object boundaries are described given a single spherical prior. Therefore, the network is tasked to estimate two radii for each ray-cast: one for the endocardial ($r_v^{in}$) and epicardial ($r_v^{out}$) boundary respectively. In practice, we frame this problem by predicting the distance to the endocardial boundary and the \textit{myocardial thickness} ($r_v^{myo}$), where $\hat{r}_v^{myo} = \hat{r}_v^{out} - \hat{r}_v^{in}$. The myocardium thickness should always be larger than 0 to obtain a topologically correct myocardium segmentation. However, this is not true for the location of the mitral and aortic valve, where the network should predict a thickness of 0. For balanced optimization, we calculate the error for $\hat{r}_v^{myo}$ as a weighted average of regions where the myocardium wall is present or not~\citep{van2023automatic}:

\begin{equation}
\mathcal{L}_{myo} = \frac{1}{M}\sum_{v \in M} |r^{myo}_v - \hat{r}^{myo}_v| + \frac{1}{K-M} \sum_{v \notin M} \hat{r}^{myo}_v,
\end{equation}

where $K$ describes the set of all vertices in a batch, and a subset of $K$ is defined that describes vertices with non-zero myocardium thickness $M = \{v: r^{myo}_v > 0\}$. Since the LV cavity is present for every vertex location, $\hat{r}^{in}_v$ is regressed using the mean absolute error:

\begin{equation}
\add{\mathcal{L}_{lv} = \frac{1}{K}\sum_{v \in K} |r_v^{in} - \hat{r}_v^{in}|.}
\end{equation}

\add{Finally, the total loss includes an additional term that regularizes the sum of $\hat{r}^{in}_v$ and $\hat{r}^{myo}_v$ to equal $r^{out}_v$ for each vertex in~$K$}:

\begin{equation}
\add{\mathcal{L}_{lv + myo} = \mathcal{L}_{lv} + \mathcal{L}_{myo} + \lambda\frac{1}{K}\sum_{v \in K} |r^{out}_v - (\hat{r}^{in}_v + \hat{r}^{myo}_v)|.}
\end{equation}

All other optimization settings are as described in Appendix~\customref{sec:gcn details}{A.3}.

\textit{Data:} A combination of Set~1 and Set~3 was used for LV cavity and LV myocardium segmentation. All \add{one-shot cross-validation} data was derived from the remaining 12 patients of Set~1. For each of these patients, the end-diastolic CCTA scan was retrospectively collected. Voxelwise reference segmentations of all cardiac chambers and the left ventricular myocardium for these patients were obtained semi-automatically with subsequent voxelwise correction as described by~\cite{bruns2022deep}. All scans from Set~3 were used as an external test set, for which whole heart reference segmentations were obtained as described by~\cite{zhuang2019evaluation}.

\begin{figure*}[b]
    \centering
    \includegraphics[trim={1cm 0.25cm 0.25cm 1.8cm}, clip, width=1.\textwidth]{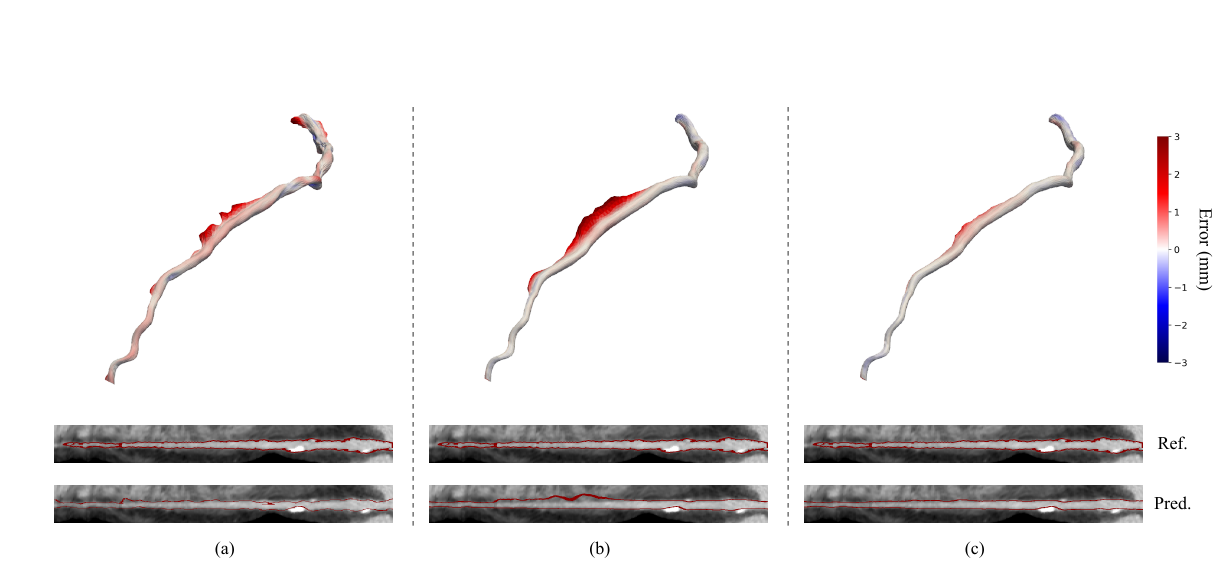}
    \caption{Meshing results for the left anterior descending artery of an image from the ASOCA dataset (Diseased 17), using different methods trained with a single annotated centerline. (a) is the output produced by a baseline GAT network, and (b) and (c) describe different versions of the proposed GEM-UNet, without and with centerline prior respectively. For each method the mesh prediction error is presented, as well as the reference and predicted lumen segmentation in a longitudinal cross-section of the MPR volume. Note how incorporating the centerline prior in (c) prevents the lumen segmentation from leaking into the right ventricle. }
    \label{fig:coronary}
\end{figure*}

\begin{table*}[t]
\centering
\caption{Lumen segmentation performance metrics. All models are trained \add{to predict reference radii from} either a single centerline (CL) or \add{all centerlines from a} single patient image (Pat), and evaluated on all other scans with available reference segmentations. \add{To evaluate performance, per-centerline meshes are warped and voxelized in 3D CCTA space to generate segmentation masks for evaluation against reference annotations.} Scores are averaged for five different training runs. A tilde~($\sim$) indicates the use of a coronary artery centerline to generate the prior mesh, while a cylindrical prior is used in all other cases. The bottom row represents an upper bound in terms of the best-performing algorithm on the ASOCA challenge test-set. Results are presented as mean (standard deviation). DSC: Dice similarity coefficient, ASSD: average symmetric surface distance, HD95: 95\% Hausdorff distance.}
\label{tab:lumen_performance}
\resizebox{0.8\textwidth}{!}{
\begin{tabular}{llcccccc}
\toprule
& $\sim$ & \multicolumn{2}{c}{DSC $\uparrow$} & \multicolumn{2}{c}{ASSD $\downarrow$ (mm)} & \multicolumn{2}{c}{HD95 $\downarrow$ (mm)} \\
\cmidrule(lr){3-4} \cmidrule(lr){5-6} \cmidrule(lr){7-8}
 & & CL & Pat & CL & Pat & CL & Pat \\
\midrule
\href{https://ieeexplore.ieee.org/abstract/document/10288599}{FanCNN} & & 0.807 (0.054) & 0.837 (0.049) & 0.752 (0.351) & 0.649 (0.334) & 1.561 (1.249) & 1.357 (1.290) \\
\href{https://link.springer.com/chapter/10.1007/978-3-030-35817-4_8}{GAT} & & 0.818 (0.047) & 0.825 (0.055) & 0.651 (0.226) & 0.649 (0.260) & 1.136 (0.882) & 1.215 (0.973) \\
GEM-CNN & & 0.807 (0.053) & 0.842 (0.045) & 0.742 (0.369) & 0.601 (0.243) & 1.516 (1.428) & 1.204 (0.953) \\
GEM-CNN & \checkmark & 0.848 (0.028) & 0.841 (0.040) & \textbf{0.567} (0.209) & 0.585 (0.218) & 1.091 (0.894) & 1.123 (0.898) \\
GEM-UNet & & 0.841 (0.038) & 0.843 (0.047) & 0.593 (0.221) & 0.587 (0.234) & 1.118 (0.898) & 1.149 (0.917) \\
GEM-UNet & \checkmark & \textbf{0.850} (0.022) & \textbf{0.852} (0.031) & \textbf{0.567} (0.210) & \textbf{0.557} (0.211) & \textbf{1.053} (0.894) & \textbf{1.090} (0.904) \\
\midrule
\href{https://asoca.grand-challenge.org/evaluation/ed9bc28a-f641-4955-b630-7ce078d1054d/}{[1]} & & \multicolumn{2}{c}{0.894 (0.034)} & \multicolumn{2}{c}{-} & \multicolumn{2}{c}{1.888 (3.365)}\\
\bottomrule
\end{tabular}
}
\end{table*}

\subsection{Tubular prior}
For mesh prediction relying on a tubular prior, the multiresolution GCN operates at 4, 2, and 1 subdivision steps respectively (see Fig.~\ref{fig:scales}, middle and bottom row). \add{Rays are cast from a template with a radius of 6.4mm to ensure coverage of a typical proximal coronary artery diameter.} The prior is tested for the task of coronary lumen segmentation, which is a prerequisite for downstream stenosis severity quantification.

\begin{table*}[t]
\centering
\caption{\add{Test-set mesh quality metrics for one-shot pericardium (Peri), LV myocardium (Myo), and coronary lumen (Lum) meshing. Marching cubes (MC) results are derived from one-shot nnUNet-based segmentations, while MC + R indicates additional remeshing post-processing. The proposed method implicitly renders a surface mesh. For each mesh, the quality metric is aggregated over five training runs. Superscript $r$ indicates a right-angled and $e$ indicates an equilateral triangulation scheme. We report the minimum angle, aspect ratio, normal consistency, Earth Mover's Distance (EMD), and self-intersections (count per mesh). Results are presented as mean~(standard deviation).}}
\label{tab:mesh_quality}
\resizebox{\textwidth}{!}{
\addtable{\begin{tabular}{llcrcccccc}
\toprule
& & \multicolumn{2}{c}{Minimum Angle $\uparrow$ (°)} & \multicolumn{2}{c}{Aspect Ratio $\downarrow$} & \multicolumn{2}{c}{Normal Consistency $\uparrow$} & EMD $\downarrow$ & Self-intersections $\downarrow$  \\
\cmidrule(lr){3-4} \cmidrule(lr){5-6} \cmidrule(lr){7-8}
&  & Mean & \multicolumn{1}{c}{Min} & Mean & Max & Mean & Min\\
\midrule
\multirow{3}{*}{Peri} & MC & 39.91 (1.68) & 29.58 (1.54) & 1.39 (0.03) & 1.59 (0.04) & 0.31 (0.01) & 0.00 (0.00) & 0.52 (0.01) & \textbf{0.00} (0.00) \\
& MC + R                  & 52.78 (0.20) & 22.21 (3.53) & 1.10 (0.00) & 2.69 
 (0.84) & 0.98 (0.00) & 0.09 (0.07) & 0.14 (0.00) & 0.03 (0.14) \\
& proposed                & \textbf{53.91} (0.29) & \textbf{32.96} (4.13) & \textbf{1.09} (0.00) & \textbf{1.56} 
 (0.18) & \textbf{0.99} (0.00) & \textbf{0.48} (0.09) & \textbf{0.13} (0.00) & \textbf{0.00} (0.00) \\
\midrule
\multirow{3}{*}{Myo} & MC  & 41.22 (2.16) & \textbf{28.65} (2.23) & 1.37 (0.04) & \textbf{1.82} (0.14) & 0.29 (0.02) & 0.00 (0.00) & 0.52 (0.01) & \textbf{0.00} (0.00) \\
& MC + R                  & 50.97 (1.75) &  4.63 (3.85) & 1.17 (0.08) & 39.09 (41.29) & \textbf{0.95} (0.03) & 0.00 (0.00) & 0.15 (0.01) & 45.72 (59.57) \\
& proposed                & \textbf{52.39} (1.03) &  7.86 (3.72) & \textbf{1.13} (0.02) &  6.00 
 (1.71) & \textbf{0.95} (0.01) & 0.00 (0.00) & \textbf{0.14} (0.00) &  \textbf{0.00} (0.00) \\
\midrule
\multirow{2}{*}{Lum} & proposed$^{r}$ & 34.81 (1.82) &  9.46 (4.83) & 1.63 (0.14) & 21.37 (54.23) & \textbf{0.96} (0.01) & 0.14 (0.15) & 0.32 (0.01) & 0.48 (2.63) \\
& proposed$^{e}$                      & \textbf{40.91} (3.20) & \textbf{13.47} (6.86) & \textbf{1.37} (0.14) & \textbf{10.32} (20.54) & \textbf{0.96} (0.01) & \textbf{0.18} (0.17) & \textbf{0.17} (0.02) & \textbf{0.18} (1.27) \\
\bottomrule
\end{tabular}}
}
\end{table*}

\subsubsection{Coronary lumen segmentation}
\label{sec: exp lumen}
Since lumen segmentation relies on a coronary artery centerline prior, the tubular geometric prior may be enforced either as a perfect cylinder or as a tube curving in 3D space, defined by the coordinates of centerline points (see Section \ref{sec:method priors}). Both of these options are explored and tested in terms of coronary artery segmentation performance. At the highest resolution, the tubular mesh is spaced at 0.5 mm between subsequent centerline points, and features 48 equiangularly distributed vertices for each centerline point. Lumen radii are regressed using the mean squared error between the predicted and reference radius for each ray-cast:

\begin{equation}
\add{\mathcal{L}_{lum} = \frac{1}{K}\sum_{v \in K} (r^l_v - \hat{r}^l_v)^2,}
\end{equation}

\add{where K is the set of all vertices in a batch, $r^l_v$ is the reference lumen radius and $\hat{r}^l_v$ is the predicted lumen radius for vertex $v$.} Additional model details and optimization settings are provided in Appendix~\customref{sec:gcn details}{A.3}.

\textit{Data:} The ASOCA dataset (Set 4) was used for the task of automatic coronary artery lumen segmentation~\citep{gharleghi2023annotated}. The 40 available training scans feature both voxelwise consensus annotations of three experts and corresponding coronary artery centerlines. An additional 20 scans were made available without annotations, for which automatically obtained segmentations may be submitted to the \href{https://asoca.grand-challenge.org/}{challenge leaderboard} for testing. For all scans, coronary artery centerlines were automatically extracted by retraining a previously published coronary artery centerline tracking algorithm using the centerlines made available in the ASOCA training dataset~\citep{wolterink2019coronary}.

\subsection{Evaluation}
The performance of each method is quantified in terms of segmentation agreement with reference annotations. For all experiments, results are evaluated in terms of 95\% Hausdorff distance (HD95), the average symmetric surface distance (ASSD), and Dice similarity coefficient (DSC). \add{Mesh quality is evaluated through multiple geometric metrics~\citep{pebay2003analysis}: minimum angle and aspect ratio quantify triangle shape quality, face normal consistency measures surface smoothness, and the number of mesh self-intersections indicates topological correctness. Additionally, the Earth Mover's Distance (EMD) is provided, which measures the amount of work necessary to transform the distribution of the scaled Jacobians of the mesh elements to a theoretically perfect distribution of all ones.}

\subsection{Experimental setting}
For segmentation of the intra-pericardial space, a single pre- and post-operative image pair is used for training, with performance tested on a hold-out test set. Similarly, a single training image is used for segmentation of the nested LV cavity and LV myocardium, for which performance is evaluated on the MM-WHS training set, which serves as our external test set. \add{At test-time, seed points are automatically identified using a landmark detection network~\citep{noothout2020deep} trained on Sets 1 and 2 to predict the center of mass of the pericardium and LV~myocardium.}

Lumen segmentation performance is tested given either a single available annotated centerline, or \add{all centerlines from a single patient} from the training set. \add{To evaluate performance, per-centerline meshes are warped and voxelized in 3D CCTA space to generate segmentation masks for comparison against reference annotations.} Performance is evaluated on all remaining training scans, using the centerlines provided by the challenge organizers.

\subsection{Baseline methods}
For all experiments utilizing a spherical prior, we include a baseline comparison with nnUNet, the current state-of-the-art segmentation network~\citep{isensee2021nnu}. \add{We further compare against two additional UNet variants: nnUNet with residual encoder layers~\citep{isensee2024nnu} and MedNeXt, a transformer-inspired architecture~\citep{roy2023mednext}. For the nested LV cavity and LV myocardium segmentation tasks, the nnUNet variants employed region-based training, which introduces a similar ordinal bias as proposed in Section~\ref{sec: exp myocardium}.} Additionally, we use FanCNN as a baseline for lumen segmentation~\citep{van2023automatic}. Similar to our proposed method, all baselines are trained in a one-shot setting.

To assess the quality of implicitly generated meshes, we compare results to a conventional marching cubes post-processing~\citep{lorensen1998marching} of the nnUNet segmentations for the intra-pericardial space and the LV myocardium. \add{We additionally compare against remeshed versions of marching cubes outputs using the Geogram library, which employs iterative mesh optimization to improve mesh quality while preserving the overall surface geometry~\citep{levy2017geogram}. Hyperparameter details on remeshing are listed in Appendix~\customref{sec:remeshing details}{A.4}.}

We further conduct several ablation studies. To assess the benefit of spherical pre-training, we compare MAE embeddings against a 1D CNN ray-cast. The effectiveness of GEM-CNN layers is evaluated by comparing them to GAT layers~\citep{wolterink2019graph}. Lastly, we compare the GEM-UNet against a GEM-CNN that operates solely on the highest mesh resolution.

\begin{figure*}[b!]
    \centering
    \includegraphics[trim={0.5cm 0.1cm 0cm 0.8cm}, clip, width=0.83\textwidth]{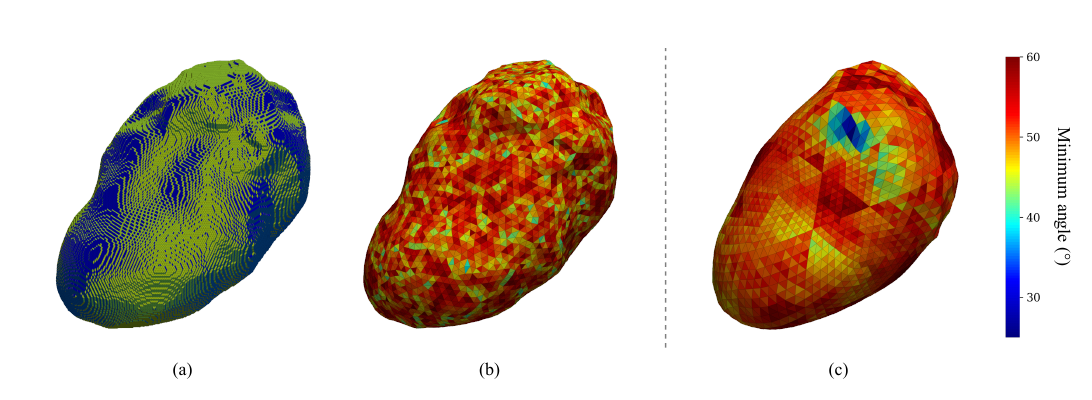}
    \caption{\add{Pericardium mesh quality comparison for a test-set image, quantified by minimum angle per face. (a) shows an nnUNet one-shot segmentation (DSC~=~0.952) for which a surface mesh is extracted using marching cubes, (b) presents a remeshed version of (a) with improved mesh quality, and (c) displays the proposed GEM-UNet segmentation (DSC = 0.960), implicitly rendering a surface mesh. Higher values indicate more equiangular triangles.}}
    \label{fig:min_angle}
\end{figure*}

\section{Results}
\label{sec:res}

\subsection{Segmentation quality}

\subsubsection{Intra-pericardial Space}
Fig.~\ref{fig:peri_dsc} displays the segmentation performance for nnUNet and four graph-based methods operating on the spherical prior, with various training subject IDs used to assess one-shot performance. In general, a steady improvement in performance can be observed given a more expressive message passing operator or network, with GAT being the least and GEM-UNet the most expressive. The GEM-UNet with MAE embeddings shows the most consistent performance regardless of the training sample, as indicated by the averaged results. It should be noted that although nnUNet performance varies significantly across training samples, it does achieve the highest peak performance. Table~\ref{tab:peri_performance} lists all averaged performance metrics for the presented methods.

\begin{figure*}[b!]
    \centering
    \includegraphics[trim={0cm 0cm 0cm 0cm}, width=0.81\textwidth]{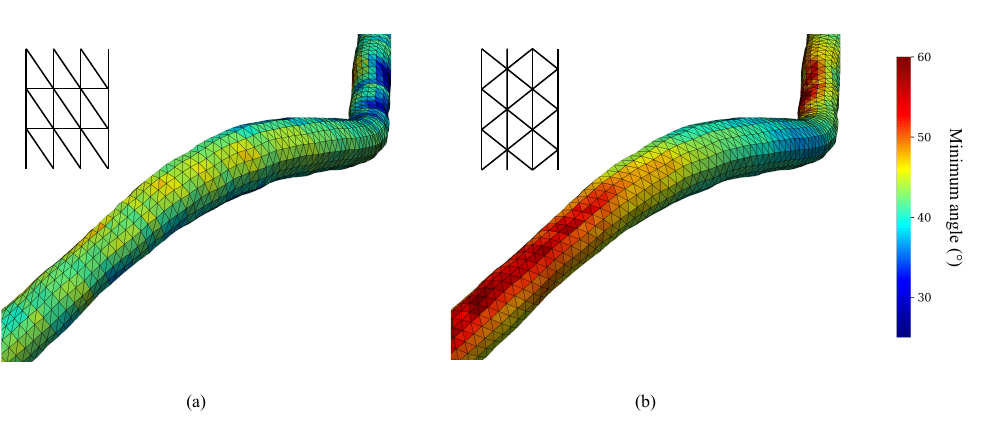}
    \caption{Mesh quality for different options of tubular surface mesh triangulation, displayed for the same coronary artery. The scheme in (a) depicts implicit right-angled triangulation, while the alternative scheme in (b) produces more equilateral triangles, potentially leading to higher mesh quality. Higher values indicate more equiangular triangles.}
    \label{fig:triangulation}
\end{figure*}

\subsubsection{Left ventricle cavity and myocardium}
Results on segmentation of the LV and LV myocardium are presented in Table~\ref{tab:model_performance}.  For all graph-based methods, utilizing MAE embeddings significantly boosts performance compared to a 1D-convolutional encoder. Interestingly, the GEM-CNN outperforms GEM-UNet in all scenarios. This difference is most notable for segmentation of the LV myocardium, where GEM-CNN with MAE embeddings achieves a DSC of 0.86 compared to 0.80 for GEM-UNet with MAE. The segmentation performance of nnUNet is similar to GEM-CNN with MAE embeddings in terms of DSC (0.89 vs 0.91 for LV, 0.85 vs 0.86 for LV myocardium), but tends to perform worse in terms of ASSD and HD95. An example of myocardium segmentation for different graph-based methods is presented in Fig.~\ref{fig:myo_comparison}.

\subsubsection{Coronary lumen}
Table \ref{tab:lumen_performance} presents the lumen segmentation performance metrics for various models. All methods were evaluated in a one-shot learning setting, using either a single centerline or a single patient image for training. The results show that the GEM-UNet with a centerline-based prior consistently outperforms other methods across all metrics. Notably, the use of a centerline prior generally improves performance compared to the cylindrical prior. The multiresolution approach (GEM-UNet) shows consistent improvements over the single-resolution GEM-CNN, highlighting the benefits of incorporating multi-scale information. A qualitative comparison of lumen segmentation results is demonstrated in Fig.~\ref{fig:coronary}.

\add{Additional qualitative results showing the worst, median, and best segmentation results are provided in Appendix~\customref{sec:quali_results}{C}, Fig.~\ref{fig:worst_median_best}.}

\subsection{Mesh quality}

\add{A comparison of mesh quality for the pericardium and LV myocardium is presented in Table~\ref{tab:mesh_quality}. For pericardium meshes, the proposed method outperforms both standard marching cubes and remeshed marching cubes across all metrics. It achieves the best mean minimum angle (53.91°), lowest aspect ratio (1.09), and best normal consistency (0.99), while completely avoiding self-intersections. For myocardium meshes, standard marching cubes shows better minimum angles (28.65° vs 7.86°) but worse performance on overall mesh quality metrics. The proposed method and remeshed marching cubes achieve similar normal consistency (0.95), but the proposed method achieves better aspect ratios (max 6.00 vs 39.09) and eliminates self-intersections, which persist even after remeshing. A qualitative comparison of meshes generated through marching cubes, including remeshing, and the proposed method is provided in Fig.~\ref{fig:min_angle}.}

Table~\ref{tab:mesh_quality} further lists coronary lumen mesh quality. The proposed method is evaluated using both the standard right-angled triangulation and an equiangular triangulation scheme (see Fig.~\ref{fig:triangulation}). Since baselines utilize the same underlying mesh prior for coronary lumen segmentation, mesh quality metrics are only presented for these two variations of the proposed method.

\subsection{\add{Computational complexity}}

\add{A detailed analysis of the computational requirements for different components of our method is provided in Appendix~\customref{sec:complexity}{B}, tables~\ref{tab:complexity_analysis1} and~\ref{tab:complexity_analysis2}. This includes CPU and GPU runtimes as well as multiply-accumulate (MAC) operations across varying mesh resolutions.}

\section{Discussion and conclusion}
\label{sec:disc}
In this work, we have presented a data-efficient approach for direct 3D surface meshing of anatomical structures with the help of shape priors. We show how different priors can be embedded for various objects of interest, and introduce a MAE pertaining strategy for 3D spherical data. We show how the proposed method produces accurate segmentations in the low-data regime while implicitly rendering manifold surface meshes.

The proposed one-shot GEM-CNN and GEM-UNet models generally outperform the nnUNet baseline in segmentation tasks for the pericardium, LV cavity, and LV myocardium. A key performance advantage of our approach is the use of a mesh prior, which implicitly guarantees genus-zero object representations. This ensures that the resulting segmentations are free from topological defects such as holes or unwanted additional segments, a property not inherently guaranteed by voxel-wise segmentation methods. While nnUNet remains a strong baseline and shows good generalization in certain cases, as seen with e.g. PVI-178 in Fig.~\ref{fig:peri_dsc}, our method offers consistent topological correctness and improved performance across the evaluated cardiac structures. The higher peak performance of nnUNet may however indicate that the icosphere subdivision level may be a potential bottleneck, which may be addressed by applying the method at higher subdivision levels. 


Tables \ref{tab:peri_performance} and \ref{tab:model_performance} further indicate that nnUNet trained on the full training dataset outperforms the proposed one-shot model. \add{In the one-shot setting, the default nnUNet configuration performs similarly to or better than both its residual encoder variant and MedNeXt, likely due to its self-configuring nature being particularly advantageous when working with limited data. Notably, MedNeXt shows reduced performance for LV cavity and myocardium segmentation, which may be a result of the method not supporting the region-based training leveraged by nnUNet.} Furthermore, it is likely that with larger amounts of training data, peak performance of the proposed method in terms of segmentation metrics is lower than that of nnUNet, as nnUNet makes fewer assumptions regarding the data and imposes a weaker inductive bias. A similar general trend may be observed when replacing CNNs with transformer architectures, for which the inductive bias is even weaker, and therefore typically boasts a higher peak performance for extremely large amounts of training data. We therefore stress the practical advantages of the proposed methods in the one-shot regime, which is specifically useful when the amount of (annotated) training data is severely limited, as is often the case in the medical image analysis domain.

The results show that the single-resolution GEM-CNN outperforms other mesh-based approaches for LV cavity and LV myocardium segmentation, including the GEM-UNet. This appears to be due to the GEM-UNet's tendency to oversmooth the myocardium thickness, as can be observed in Fig.~\ref{fig:myo_comparison}b, resulting in an oversegmentation of the myocardium wall. In contrast, the single-resolution GEM-CNN, with MAE ray-cast embeddings correctly identifies the myocardium boundary. This suggests that the multi-resolution approach, while beneficial for some structures, may introduce challenges in capturing fine details for nested object segmentation, especially when the outer radius may equal zero. The GEM-CNN seems to preserve such local features better, thus resulting in better outer wall delineations.

As indicated by the results for all segmentations with a spherical prior, the addition of MAE ray-cast embeddings boosts performance significantly (see Tables \ref{tab:peri_performance} and \ref{tab:model_performance}). Despite the apparent benefits of MAE pretraining, extensive experiments were limited to a \mbox{ViT-S} backbone with different masking percentages due to lengthy optimization times. \add{Initial attempts to directly leverage MAE embeddings for vertex-wise displacement prediction using a simple linear layer yielded suboptimal outcomes. We believe this is because, while MAE captures broad spatial awareness, the fine-grained radial regression required benefits from the locality-biased processing provided by subsequent GCN layers.  Furthermore, the inherent smoothness bias in MAE reconstructions, stemming from the mean squared error loss, may be ill-suited for predicting detailed features (see Appendix~\customref{sec:quali_results}{C}, Fig.~\ref{fig:mae_reconstruction}).} These constraints highlight the need for further investigation into optimal pretraining strategies and architectures. The transformer encoder may for example theoretically be applied at arbitrary resolutions, thanks to the continuous positional embeddings generated through spherical harmonics rather than learned embeddings. Furthermore, cylindrical harmonics may be leveraged similarly to pretrain an MAE for downstream tasks on tubular structures.

For lumen segmentation, the multiresolution GEM-UNet consistently outperforms all single-resolution methods, illustrating the advantages of incorporating multi-scale information (see Table~\ref{tab:lumen_performance}). Despite removing equivariance with respect to cylindrical symmetry, the incorporation of the coronary artery centerline as a prior consistently improves performance for both GEM-CNN and GEM-UNet, suggesting that this prior contributes to more accurate surface representations. Future work may utilize this prior for other tasks which may benefit from implicit curvature information, such as direct prediction of fractional flow reserve~\citep{tesche2020machine, hampe2022deep}.

FanCNN and GEM-CNN with cylindrical prior perform nearly identically, which may be attributed to their comparable data representation approaches, both being cylinder representations. However, unlike FanCNN, the graph representation does not require padding and achieves the same performance with a more compact representation. This efficiency extends to multiscale analysis as well; The GEM-UNet can perform such analysis without the need to conform to specific image sizes divisible by powers of two, a common constraint in traditional UNet-like architectures.

\add{An evaluation of mesh quality metrics is presented in Table~\ref{tab:mesh_quality}. The results highlight some interesting differences between marching cubes with remeshing and our implicit approach. Mesh quality metrics are generally lower for the myocardium compared to the pericardium across all methods, likely due to the more complex nested structures requiring more sudden mesh deformations. While remeshing improves mesh quality over standard marching cubes, its performance varies between anatomical structures. For the pericardium, remeshing achieves good quality metrics, but shows more pronounced quality degradation on the myocardium in terms of both aspect ratios and minimum angles. A key advantage of our method's implicit shape prior is that it guarantees meshes free from self-intersections when using a spherical prior - though this guarantee does not extend to tubular mesh generation, where highly tortuous arteries may result in self-intersections due to intersecting MPR tangent planes. For coronary arteries, mesh quality tends to be lower using the implicit scheme compared to structures with an icosphere prior, which can be attributed to the higher curvatures typically found in tubular surfaces and the initialization with right-angled triangles limiting minimum angles to 45 degrees. Future work may investigate ways to directly impose topological constraints on output meshes during training, and different mesh connectivity priors could be introduced implicitly to improve mesh quality and ensure direct translatability to flow modeling (see Fig. \ref{fig:triangulation}).}


The effectiveness of the proposed method relies on the availability of suitable coordinate priors.~\add{While the method is robust to moderate initialization variations through augmentation, extremely poor initializations, such as near the structure boundary, may lead to suboptimal results.} Seed points can however be generated relatively simply, as a spherical prior only requires a single coordinate to act on. Such points can be extracted manually, or by training a lightweight CNN to automatically identify landmarks~\citep{noothout2020deep, ao2023feature}. Similarly, centerline extraction is a common initial step for downstream visualization and analysis of vascular structures~\citep{wolterink2019coronary, alblas2023sire}. 

It should be noted that including MAE embeddings comes with the trade-off of removing SO(3) equivariance on the sphere. Though this is not typically an issue, which can further be accounted for by randomly rotating the spherical harmonics during training, it does remove a useful implicit feature of GEM convolutions. Future work may investigate equivariant transformers for more expressive data representations~\citep{fuchs2020se, liao2023equiformerv2}.

In conclusion, we present a flexible and data-efficient method for direct 3D surface meshing of anatomical structures that effectively incorporates geometric priors. Our approach demonstrates good performance in low training data regimes while consistently producing topologically correct, manifold surface meshes across various cardiac structures and coronary arteries.

\section*{Acknowledgments}
This study was supported by the DLMedIA program \mbox{(P15-26)} funded by Dutch Technology Foundation with participation of Philips Healthcare. This study was also partially funded by a private-public partnership grant provided by Health Holland, with contributions from B. Braun Melsungen AG, Germany, and Infraredx, Inc., Bedford, MA, USA (DEBuT-LRP TKI-PPP, grant no. NCT04765956).

\bibliographystyle{model2-names.bst}\biboptions{authoryear}
\bibliography{medima-template}

\clearpage

\section*{Supplementary Material}
\section*{A. Implementation details}

\subsection*{A.1 Convolutional encoder}
\label{sec:conv encoder}
The implementation details for the 1D convolutional encoder are provided in Table~\ref{tab:conv encoder}. Each convolution block consists of two regular convolutions followed by a 2-strided convolution, of which the output is added to the downsampled input of the block.

\begin{table}[h]
\centering
\resizebox{0.8\columnwidth}{!}{
\begin{tabular}{l|l|l}
 & \textbf{config} & \textbf{value} \\
\hline
\multirow{4}{*}{\begin{tabular}[c]{@{}l@{}}Pericardium\\[0.1em] Myocardium\end{tabular}} 
 & $n_v$ & 256 \\
 & spacing & 0.5 mm \\
 & conv blocks & [8, 8, 16, 16, 32, 32, 64] \\
 & params & 55K\\
\hline
\multirow{4}{*}{Lumen} 
 & $n_v$ & 32 \\
 & spacing & 0.2 mm \\
 & conv blocks & [8, 8, 16] \\
 & params & 3K\\

\end{tabular}}
\caption{Implementation details for the convolutional encoder.}
\label{tab:conv encoder}
\end{table}

\subsection*{A.2 Masked autoencoder}
\label{sec:masked autoencoder}
Implementation details for spherical autoencoder pre-training. Configuration for the Encoder and Decoder are given, as well as training details. Positional embeddings are provided through spherical harmonics up until the 8th degree. Most optimization details were derived from the original MAE paper~\citep{he2022masked}, the largest difference being the masking percentage. Initial experiments showed that even when masking a high percentage of ray-casts, the method was able to reconstruct the full icosphere with relative ease.

\begin{table}[h]
\centering
\resizebox{0.9\columnwidth}{!}{
\begin{tabular}{l|l|l}
 & \textbf{config} & \textbf{value} \\
 \hline
& spherical harmonics degree & 1 to 8 \\
\hline
\multirow{5}{*}{Encoder (ViT-S)} 
 & layers & 12 \\
 & width & 384 \\
 & heads & 6 \\
 & head width & 64 \\
 & params & 21.4M \\
\hline
\multirow{5}{*}{Decoder}
 & layers & 4 \\
 & width & 64 \\
 & heads & 6 \\
 & head width & 64 \\
 & params & 527K \\
 \hline
\multirow{10}{*}{Optimization}
& optimizer & AdamW \\
& base learning rate & 1.5e-4\\
& weight decay & 0.05\\
& optimizer momentum & $\beta_1$, $\beta_2$ = 0.9, 0.95\\
& batch size & 16\\
& learning rate schedule & cosine decay\\
& scheduler hyperparameters & $T_0$, $T_{mult}$ = 40, 2\\
& scheduler cycles & 7\\
& warmup epochs & 40\\
& masking percentage & 95\%\\
\hline
\multirow{3}{*}{Augmentation}
& RandomDisplacement & [-40, 40] mm \\
& RandomRotate & any \\
& RandomScale & [0.7, 1.4]
\end{tabular}}
\caption{Implementation details for the masked autoencoder.}
\label{tab:masked autoencoder}
\end{table}

\newpage
\subsection*{A.3 GCN}
\label{sec:gcn details}
Architecture and training details for the different graph-based approaches.

\begin{table}[h]
\centering
\resizebox{0.9\columnwidth}{!}{
\begin{tabular}{l|l|l}
 & \textbf{config} & \textbf{value} \\
 \hline
\multirow{4}{*}{GAT} 
 & layers & 3 \\
 & heads & 4 \\
 & head width & 32 \\
 & params & 84.6K \\
 \hline
\multirow{3}{*}{GEM-CNN} 
 & layers & 3 \\
 & width & 16 \\
 & params & 86.2K \\
  \hline
\multirow{3}{*}{GEM-UNet} 
 & depth layers & 3 \\
 & conv blocks & [16, 16] \\
 & params & 326K \\
  \hline
 \multirow{8}{*}{Optimization}
& optimizer & AdamW \\
& base learning rate & 0.01\\
& optimizer momentum & $\beta_1$, $\beta_2$ = 0.9, 0.95\\
& batch size & 10\\
& learning rate schedule & cosine decay\\
& scheduler hyperparameters & $T_0$, $T_{mult}$ = 90, 2\\
& scheduler cycles & 2\\
& warmup epochs & 10\\
\hline
\multirow{3}{*}{\begin{tabular}[c]{@{}l@{}}Augmentation\\[0.1em] (Pericardium)\end{tabular}} 
& RandomDisplacement & [-40, 40] mm \\
& RandomRotate & any \\
& RandomScale & [0.7, 1.4] \\
\hline
\multirow{3}{*}{\begin{tabular}[c]{@{}l@{}}Augmentation\\[0.1em] (Myocardium)\end{tabular}} 
& RandomDisplacement & [-10, 10] mm \\
& RandomRotate & any \\
& RandomScale & [0.7, 1.4]
\end{tabular}}
\caption{Implementation details for the various graph convolutional networks.}
\label{tab:masked autoencoder}
\end{table}

\subsection*{\add{A.4 Remeshing}}
\label{sec:remeshing details}
\add{Here, we list the hyperparameter details used in the Geogram remeshing library. The number of vertices is chosen to be approximately double the resolution of our proposed method, while repair flags and hole area constraints help maintain mesh integrity when processing potentially imperfect segmentations.}

\begin{table}[h]
\centering
\resizebox{0.65\columnwidth}{!}{
\addtable{\begin{tabular}{l|l|l}
& \textbf{config} & \textbf{value} \\
\hline
\multirow{2}{*}{Vertices}
& Pericardium & 5124 \\
& Myocardium & 10248 \\
\hline
\multirow{5}{*}{Remeshing}
& anisotropy & 1 \\
& pre-repair & true \\
& post-repair & true \\
& max. hole area & 100 \\
& remove internal shells & true \\
\end{tabular}}}
\caption{\add{Implementation details for geogram remeshing post-processing.}}
\label{tab:remeshing}
\end{table}

\section*{\add{B. Time-space complexity analysis}}
\label{sec:complexity}

\add{An analysis of the time and space complexity of the models presented in this paper is presented in Tables~\ref{tab:complexity_analysis1} and~\ref{tab:complexity_analysis2}. We list both CPU and GPU runtime for different parts of the models, as well as the number of MAC operations, with all results averaged over 100 runs. Scaleability is evaluated by running the model at higher mesh resolutions, which is achieved by further subdivision of the proposed priors. All GPU-based experiments were conducted on an NVIDIA RTX~2080~Ti.}

\begin{table*}[t]
\centering
\caption{\add{Space-time complexity analysis for ray-casting and different ray-cast encoders. Time measurements (in seconds) are reported for both CPU and GPU implementations across different mesh resolutions. Parameters and MACs (multiply-accumulate operations) indicate model complexity. Models are evaluated at several mesh resolutions, indicated by the prior subdivision level.}}
\label{tab:complexity_analysis1}
\resizebox{0.95\textwidth}{!}{
\addtable{\begin{tabular}{lccccccccc}
\toprule
& & \multicolumn{2}{c}{Ray-casting} & \multicolumn{3}{c}{ConvEncoder (\#P: 55.2K / 3.11K)} & \multicolumn{3}{c}{MAE (\#P: 21.4M)} \\
\cmidrule(lr){3-4} \cmidrule(lr){5-7} \cmidrule(lr){8-10}
Prior & Vertices & CPU (s) & GPU (s) & CPU (s) & GPU (s) & MACs & CPU (s) & GPU (s) & MACs \\
\midrule
Icosphere-4 & 2.56K & \num{8.14e-2} & \num{1.46e-5} & \num{1.03e-1} & \num{1.52e-2} & 1.53G & \num{8.66e-1} & \num{5.11e-2} & 116G \\
Icosphere-5 & 10.2K & \num{2.72e-1} & \num{2.23e-5} & \num{3.21e-1} & \num{4.79e-2} & 6.14G & \num{1.67e1} & \num{7.29e-1} & 1.19T \\
Icosphere-6 & 40.9K & \num{9.61e-1} & \num{1.50e-5} & \num{1.65e0} & \num{4.69e-2} & 24.5G & - & - & - \\
\midrule
Cylinder-4 & 4.80K & - & - & \num{2.91e-3} & \num{4.80e-4} & 29.4K & - & - & - \\
Cylinder-5 & 19.2K & - & - & \num{2.31e-2} & \num{7.51e-4} & 118K & - & - & - \\
Cylinder-6 & 76.8K & - & - & \num{7.71e-2} & \num{1.98e-3} & 470K & - & - & - \\
\bottomrule
\end{tabular}}
}
\end{table*}

\begin{table*}[t]
\centering
\caption{\add{Space-time complexity analysis for graph-based decoder methods. Time measurements (in seconds) are reported for both CPU and GPU implementations across different mesh resolutions. Parameters and MACs indicate model complexity. Models are evaluated at several mesh resolutions, indicated by the prior subdivision level.}}
\label{tab:complexity_analysis2}
\resizebox{\textwidth}{!}{
\addtable{\begin{tabular}{lcccccccccc}
\toprule
& & \multicolumn{3}{c}{GAT (\#P: 84.6K)} & \multicolumn{3}{c}{GEM-CNN (\#P: 86.2K)} & \multicolumn{3}{c}{GEM-UNet (\#P: 326K)} \\
\cmidrule(lr){3-5} \cmidrule(lr){6-8} \cmidrule(lr){9-11}
Prior & Vertices & CPU (s) & GPU (s) & MACs & CPU (s) & GPU (s) & MACs & CPU (s) & GPU (s) & MACs \\
\midrule
Icosphere-4 & 2.56K & \num{6.08e-2} & \num{7.27e-3} & 3.28M & \num{1.29e-1} & \num{3.83e-2} & 12.5M & \num{2.15e-1} & \num{7.85e-2} & 22.6M \\
Icosphere-5 & 10.2K & \num{2.05e-1} & \num{1.49e-2} & 13.1M & \num{3.90e-1} & \num{5.69e-2} & 49.8M & \num{6.12e-1} & \num{9.25e-2} & 90.1M \\
Icosphere-6 & 40.9K & \num{6.35e-1} & \num{6.41e-2} & 52.5M & \num{2.18e0} & \num{1.48e-1} & 199M & \num{3.05e0} & \num{2.49e-1} & 360M \\
\midrule
Cylinder-4 & 4.80K & \num{5.50e-2} & \num{5.20e-3} & 146M & \num{1.46e-1} & \num{1.68e-2} & 160M & \num{2.45e-1} & \num{5.12e-2} & 179M \\
Cylinder-5 & 19.2K & \num{3.82e-1} & \num{1.49e-2} & 584M & \num{9.17e-1} & \num{5.41e-2} & 640M & \num{1.41e0} & \num{1.08e-1} & 716M \\
Cylinder-6 & 76.8K & \num{1.61e0} & \num{5.13e-2} & 2.34G & \num{4.36e0} & \num{2.33e-1} & 2.56G & \num{6.10e0} & \num{3.52e-1} & 2.87G \\
\bottomrule
\end{tabular}}
}
\end{table*}

\newpage
\null
\vfill

\clearpage

\add{\section*{C. Additional qualitative results}}
\label{sec:quali_results}

\begin{figure}[H]
    \centering
    \begin{minipage}{\textwidth}
    \includegraphics[trim={0cm 0cm 0cm 0cm}, width=\textwidth]{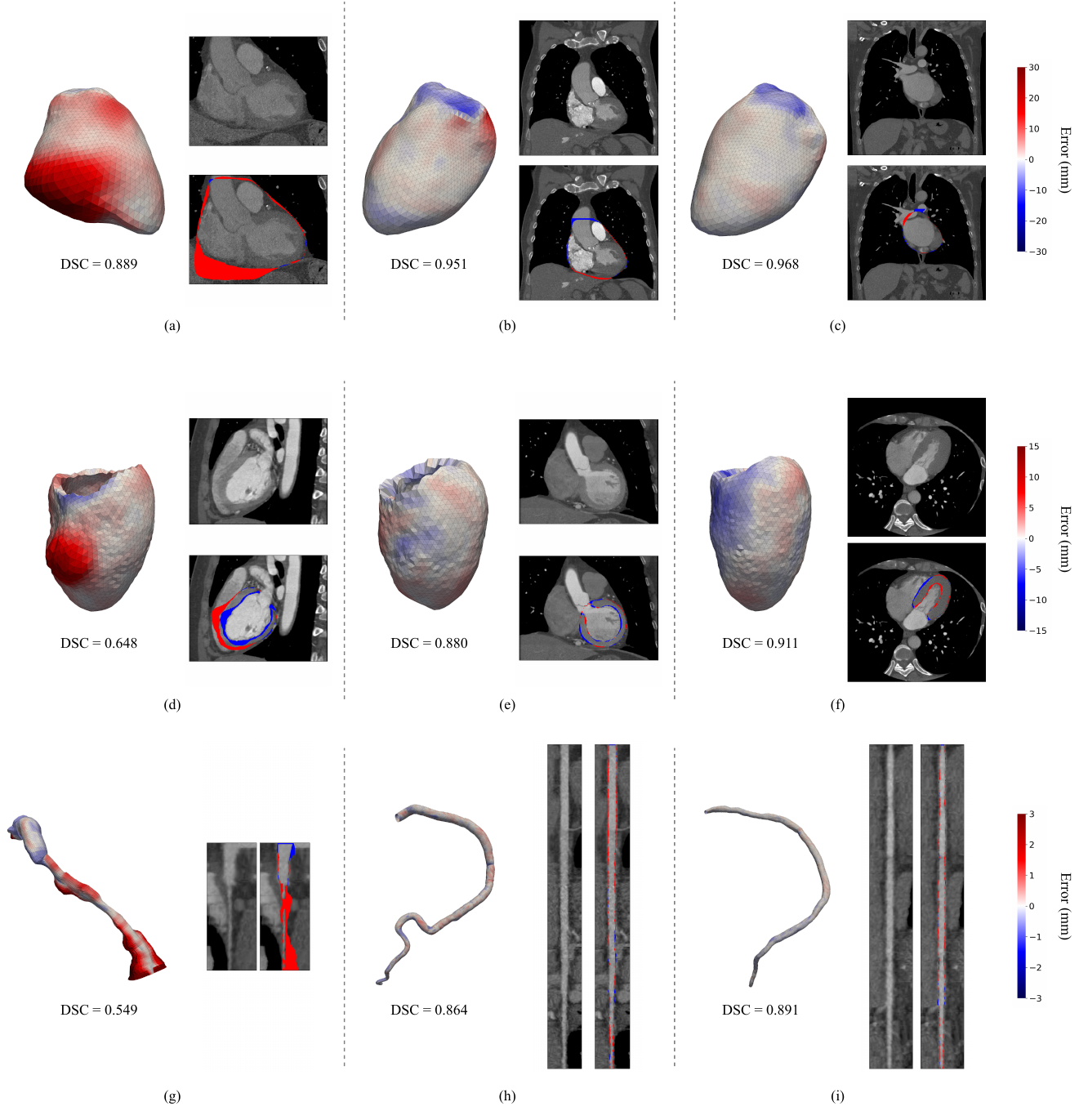}
    \caption{\add{Qualitative results showing worst (left), median (middle), and best (right) performing cases for one-shot surface meshing. Top row (a-c): Pericardium segmentation, showing the mesh with error heatmap and corresponding CCTA slices with reference contours. Middle row (d-f): LV myocardium segmentation, displaying the predicted mesh with error visualization and CCTA slices with reference annotations. Bottom row (g-i): Coronary artery lumen segmentation, presenting the mesh predictions with error mapping and corresponding multi-planar reformatted (MPR) views. For each case, the Dice similarity coefficient (DSC) is reported. Error heatmaps indicate the local deviation from the reference segmentation in millimeters, with red showing overestimation and blue showing underestimation.}}
    \label{fig:worst_median_best}
    \end{minipage}
\end{figure}

\clearpage

\begin{figure}[H]
    \centering
    \begin{minipage}{\textwidth}
    \includegraphics[trim={-1cm 0cm 0cm 0cm}, width=\textwidth]{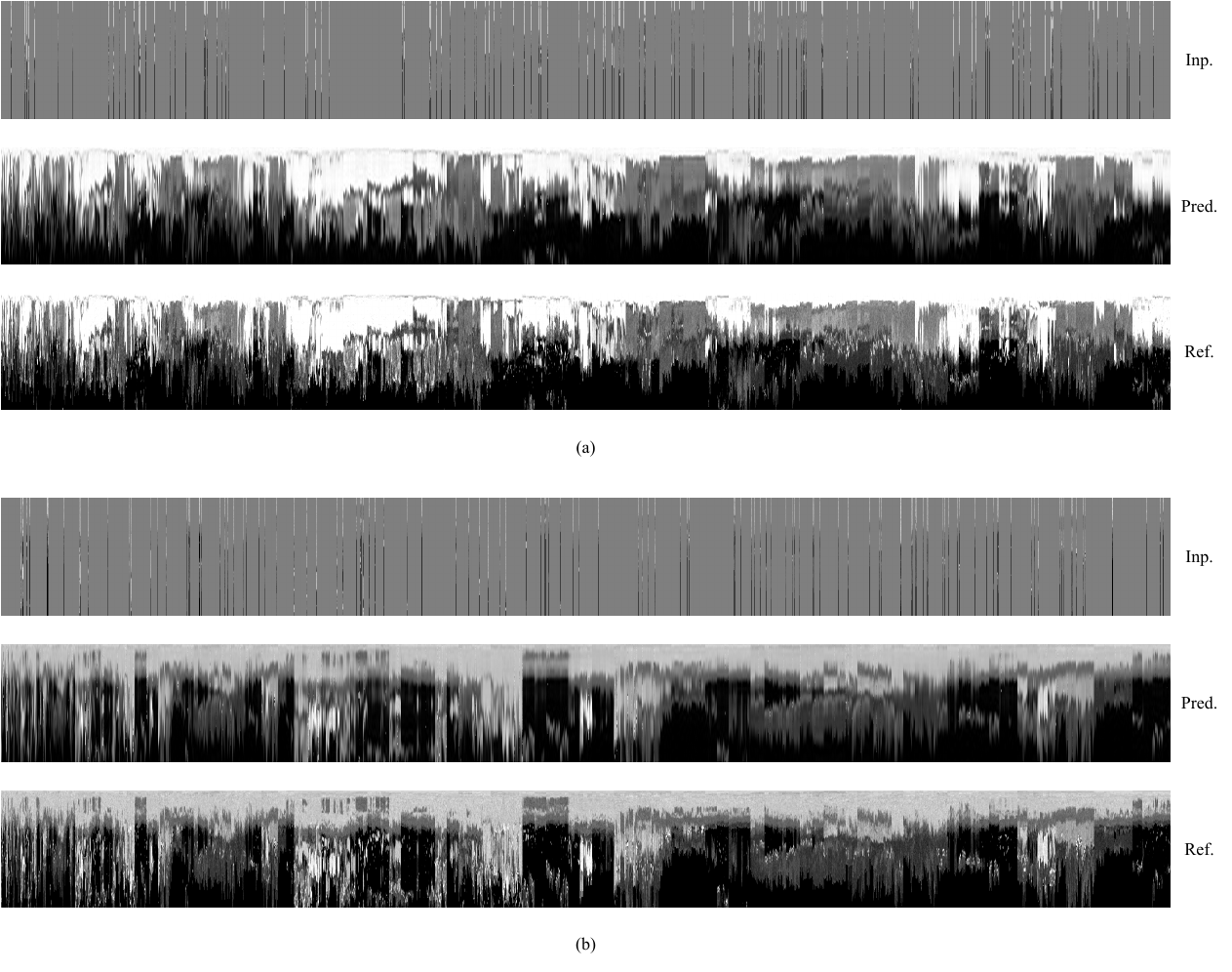}
    \caption{\add{Visualization of spherical MAE reconstruction results for ray-cast signals. Each row shows the input signal (Inp.), predicted reconstruction (Pred.), and reference (Ref.) respectively for (a) a pericardium seed point and (b) a left ventricle myocardium seed point. The input shows the masked ray-cast input signal, where most of the image information is hidden. The model reconstructs the complete signal pattern, closely matching the true image data. Each vertical line represents a ray-cast from the central seed point outward, with intensity values sampled along the ray.}}
    \label{fig:mae_reconstruction}
    \end{minipage}
\end{figure}

\end{document}